\documentclass[prl,aps,twocolumn,showpacs]{revtex4-1}

\def\beq{\nopagebreak \begin{equation}}
\def\eeq{\end{equation}}

\usepackage{graphicx}
\usepackage{amsmath}
\usepackage{caption}
\usepackage{braket}
\usepackage{color}
\definecolor{forestgreen}{RGB}{34, 139, 34}

\begin{document}
\title{
Exceptionally strong phonon scattering by B substitution in cubic SiC}

\author{Ankita Katre}
\email{ankitamkatre@gmail.com}
\author{Jes\'{u}s Carrete}
\author{Bonny Dongre}
\author{Georg K. H. Madsen}
\author{Natalio Mingo}
\email{natalio.mingo@cea.fr}
\affiliation{LITEN, CEA-Grenoble, 17 rue des Martyrs, 38054 Grenoble Cedex 9, France, \\
	    Institute of Materials Chemistry, TU Wien, A-1060 Vienna, Austria}


\begin{abstract}

We use {\it ab-initio} calculations to predict the thermal conductivity of cubic SiC with different types 
of defects. An excellent quantitative agreement with previous experimental measurements is found.
The results unveil that $\mathrm{B_{C}}$ substitution has a much stronger effect than any of the other defect types in 3C-SiC, 
including vacancies. This finding 
contradicts the prediction of the classical mass-difference model of impurity scattering, according to which the effects 
of $\mathrm{B_{C}}$ and $\mathrm{N_{C}}$ would be similar and much smaller than that of the C vacancy.
The strikingly different behavior of the $\mathrm{B_{C}}$ defect arises from a unique pattern of resonant 
phonon scattering caused by the broken structural symmetry around the $\mathrm{B}$ impurity. 
\end{abstract}

\maketitle

Silicon carbide (SiC) plays a fundamental role in many emerging technologies, ranging from biomedical sensors to optoelectronics, power electronics and photovoltaics \cite{Eddy_Sci2009, Bhatnagar_ieee1993, Frechette_jacs2006, Hao_jacs2016, Saddow_2012, Hillenbrand_Nature2002, Mikael_SEMSC2016, Beaucarne_PPRA2002, Cicero_PRL2004}.
Most notably, this material has been termed the ``linchpin to green energy''
that may replace Si-based technology in power electronics \cite{Eddy_Sci2009}, owing partly to its large lattice thermal conductivity ($\kappa$).
From the many stable polytypes of SiC \cite{Harris_1995}, 
two of the hexagonal ones, 6H-SiC and 4H-SiC, 
have been extensively studied and widely used \cite{Zheludev_2007, Lien_ieee2014, Harris_1995}. 
In contrast, the structurally less complex cubic polytype of SiC with zinc-blende structure (3C-SiC) 
is much less well understood, despite presumably having the best electronic properties \cite{Veruchhi_jacs2012}, 
and, as we will see, possibly a higher $\kappa$ than the other polytypes. This is partly due to the difficulty in synthesizing 
high quality crystals, although recent improvements in 3C-SiC growth techniques have prompted a renewed interest in it \cite{Veruchhi_jacs2012}.

Surprisingly, the 
reference measurements of $\kappa$ on pure undoped 3C-SiC are over 20 years old and little detail is known about the quality of the samples \cite{Harris_1995, Morelli_1993}. The reference value of $\kappa$ for 3C phase 
is perplexingly lower than that for the structurally more complex 6H phase, 
raising doubts about whether this is truly an intrinsic property or just a consequence of the defective, polycrystalline quality of the 3C-SiC samples.
It is then clear that to understand the conduction properties of 3C-SiC, and to harness its full potential, 
one must first
comprehend the way defects affect it. 
As we show here, 
by comparing predictive \textit{ab-initio} calculations with experiments on defective samples, 
a richer physical picture emerges, unveiling the striking differences in the way different dopants affect $\kappa$. 
This also indirectly suggests that the intrinsic $\kappa$ of defect-free 3C-SiC should be much higher 
than previously reported and surpass that of the 6H phase.

In this paper, we compare our results to the 
$\kappa(T)$ curves for doped samples of 3C-SiC \cite{Ivanova_im2006}.
We use an \textit{ab-initio} approach to quantify 
the phonon scattering rates of $\mathrm{N_{C}}$ substitutional defects. 
The predicted $\kappa$ is in excellent agreement with the experimental results. 
This then allows us to explain the effect of codoping with N and B, and shows that B impurities scatter phonons two 
orders of magnitude more strongly overall than N or Al impurities, and as strongly as C vacancies. We identify resonant phonon 
scattering as the reason for this behavior, resulting from distortion and broken symmetry around the B atom. We then 
show this resonant scattering behavior to be a general phenomenon that can show up for strong enough perturbations.


$\kappa$ for cubic structures can be obtained from a complete phonon picture 
as explained in Refs.~\onlinecite{Li_CPC2014, Katre_PRB2016}.
We use an iterative scheme to solve the full linearized Boltzmann transport equation (BTE) 
and calculate the $\kappa$ of 3C-SiC as implemented in the almaBTE code \cite{ALMA_BTE}. 
Details about the calculations are presented in the supplementary material.

\begin{figure}[t]
 \centering
  \includegraphics[width=8.5cm]{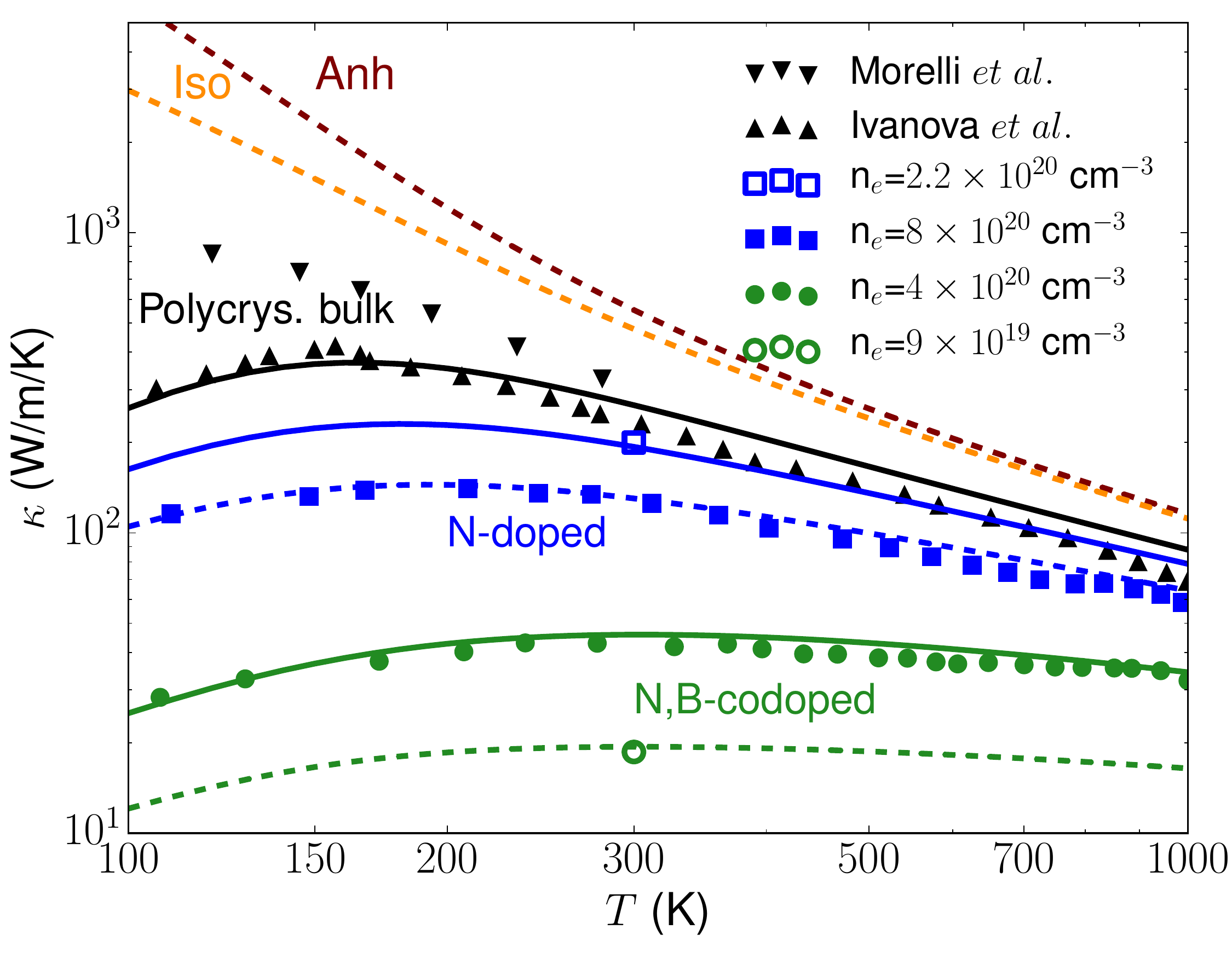} 
  \caption{Calculated $\kappa$ of 3C-SiC including respectively anharmonic phonon scattering, 
  isotope scattering, grain boundary scattering (\noindent\textcolor{black}{\rule{0.4cm}{1.5pt}} $L_{\text{grain}}=0.8~\mu$m) 
  and different concentrations of defects 
 [\noindent\textcolor{blue}{\rule{0.4cm}{1.5pt}}
   $2.2\times10^{20}$ cm$^{-3}$ N$_{\text{C}}$ (0.45$\%$), 
   \noindent\textcolor{blue}{\rule{0.15cm}{1.5pt}}\textcolor{white}{\rule{0.1cm}{1.5pt}}\textcolor{blue}{\rule{0.15cm}{1.5pt}} 
   $8\times10^{20}$ cm$^{-3}$ N$_{\text{C}}$ (1.63$\%$), 
   \noindent\textcolor{forestgreen}{\rule{0.4cm}{1.5pt}} $6.6\times10^{20}$ cm$^{-3}$ N$_{\text{C}}$ (1.3$\%$),  
   $2.6\times10^{20}$ cm$^{-3}$ B$_{\text{C}}$ (0.5$\%$), and  
   \noindent\textcolor{forestgreen}{\rule{0.15cm}{1.5pt}}\textcolor{white}{\rule{0.1cm}{1.5pt}}\textcolor{forestgreen}{\rule{0.15cm}{1.5pt}}
   $1.32\times10^{21}$ cm$^{-3}$ N$_{\text{C}}$ (2.7$\%$), $1.23\times10^{21}$ cm$^{-3}$ B$_{\text{C}}$ (2.5$\%$)]. 
   The experiments are from Morelli \textit{et al.} \cite{Morelli_1993} and Ivanova \textit{et al.} \cite{Ivanova_im2006}. 
 }
 \label{fig:kappa_l}    
\end{figure}

The calculated $\kappa$ for defect-free single-crystal 3C-SiC
is shown as the brown dashed line in Fig.~\ref{fig:kappa_l}. Its value at 300~K
is 552~W/m/K, which is about $\sim$70$\%$ higher than the experimentally reported $\kappa$ (320~W/m/K) \cite{ioffe_sic, Morelli_1993}.
It also surpasses the experimental $\kappa$ at 300~K for the 6H-SiC phase 
by $\sim$10$\%$ \cite{ioffe_sic}. 
Inclusion of Si and C isotope scattering slightly lowers the calculated $\kappa$ (orange dashed line in Fig.~\ref{fig:kappa_l}), but the values remain much larger than the experimental ones.

\begin{figure}[t]
 \centering
 \includegraphics[width=8.5cm]{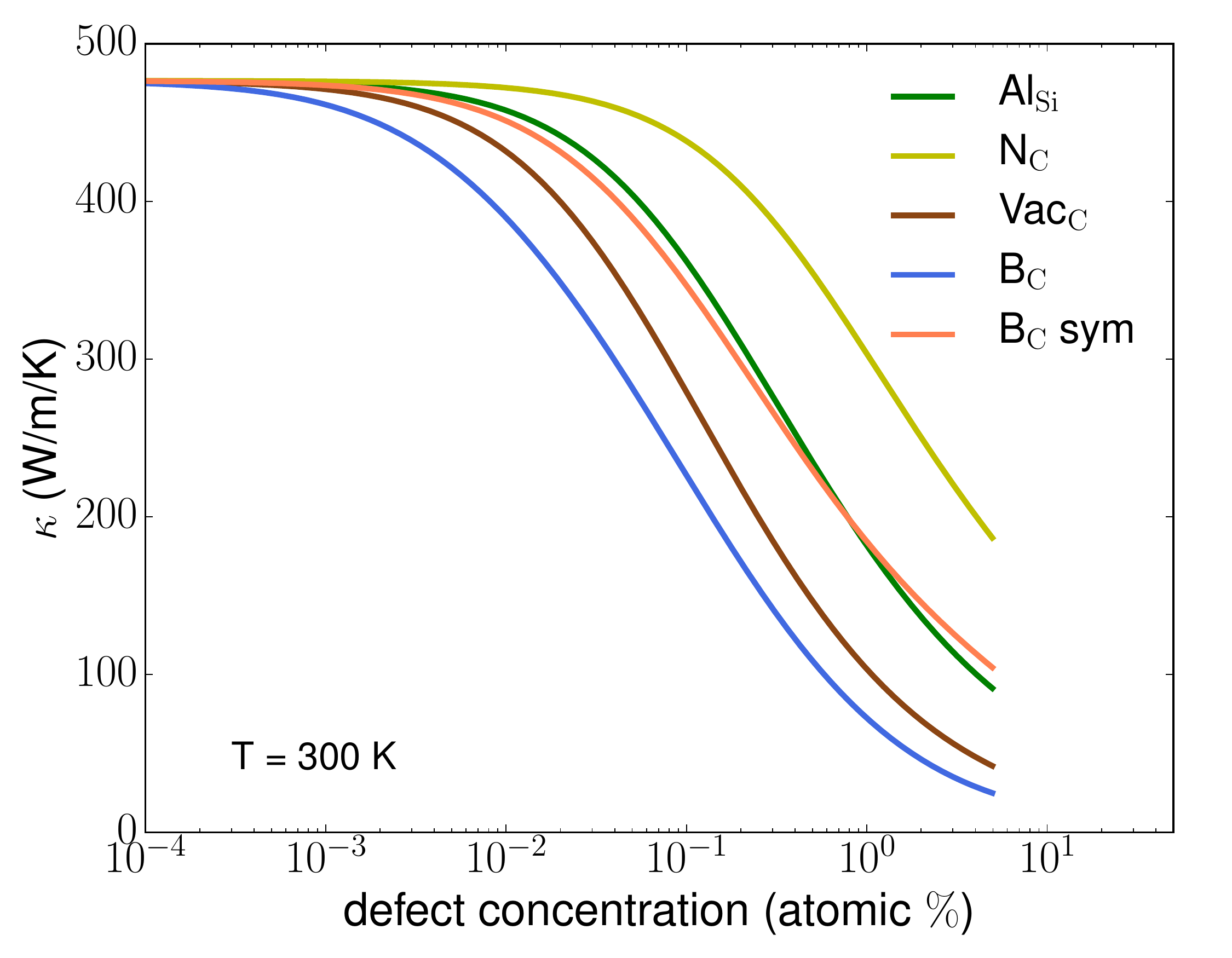} 
 \caption{3C-SiC $\kappa$ variation with defect concentration for Al$_{\text{Si}}$, 
  N$_{\text{C}}$, Vac$_{\text{C}}$ and B$_{\text{C}}$ defects.}
 \label{fig:kappaconc} 
\end{figure}

The calculated $\kappa$ for the defective structures of 
3C-SiC are compared with experiments on polycrystalline samples from Ref.~\onlinecite{Ivanova_im2006} in Fig.~\ref{fig:kappa_l}. 
The additional contribution of the grain boundaries to the scattering rate is included in the standard way as $\tau^{-1}_{\text{grain}}=v/L_{\text{grain}}$, where $v$ is 
the phonon group velocity and $L_{\text{grain}}$ is the grain size. Employing $L_{\text{grain}}=0.8~\mu$m for all the samples yields good agreement with experiment. 

The black filled triangles in Fig.~\ref{fig:kappa_l} are the experimental measurements on the polycrystalline bulk sample 
without any dopant \cite{Ivanova_im2006}. The other experimental results are for the N-doped, 
and N-and-B-codoped n-type samples in Ref.~\onlinecite{Ivanova_im2006}. 
The case of N-only doping demonstrates the predictive power of the \textit{ab-initio} approach.
As N in SiC can be considered as a shallow 
donor \cite{Gupta_2006, Smith_jem1999, Isoya_pbcm2003} and each dopant contributes one 
electron per defect atom, the concentrations  of N$_{\text{C}}$ defects can be expected 
to be the same as the carrier concentration. Using the carrier 
concentrations for different N-doped 3C-SiC samples 
reported in Ref.~\onlinecite{Ivanova_im2006} yields excellent agreement between the calculated and the experimental $\kappa$ for the whole temperature range 
in Fig.~\ref{fig:kappa_l} (blue lines and symbols). This confirms the reliability of the \textit{ab-initio} approach.

The case of B and N codoping is considered next. 
B substitutes for C in 3C-SiC when it is grown in Si-rich conditions \cite{Fukumoto_prb1996, Bechstedt_jpcm2001, Petrenko_prb2016}. 
The fact that the undoped samples in Ref.~\onlinecite{Ivanova_im2006} are $n$-type with a reasonably high carrier 
concentration ($\approx 10^{17}$cm$^{-3}$) would indicate the formation of vacancies at C sites, 
which happens only in Si-rich growth conditions \cite{Oda_JCP13}. 
Due to codoping, the carrier
concentrations given in Ref.~\onlinecite{Ivanova_im2006} correspond to the 
difference between the donor and acceptor concentrations, and thus they do not 
uniquely determine the defect densities. In this case, the \textit{ab-initio} calculations are able to determine individual defect densities: by keeping the difference in concentrations equal 
to the experimental carrier concentration, the calculation matches the whole temperature 
dependent experimental curve only if the individual concentrations are 1.3$\%$ and 
0.5$\%$ for N$_{\text{C}}$ and B$_{\text{C}}$ respectively (Fig.~\ref{fig:kappa_l}: upper green line and circles). 
The relative concentrations of the other codoped sample were similarly derived 
yielding 2.7$\%$ and 2.5$\%$ of N$_{\text{C}}$ and B$_{\text{C}}$ defects 
respectively (lower green line and circle in Fig.~\ref{fig:kappa_l}). 
Our calculations capture the correct experimental trend of a
big reduction in $\kappa$ seen when B$_{\text{C}}$ defects, even in smaller concentrations 
than N$_{\text{C}}$, are introduced in the samples (Fig.~\ref{fig:kappa_l}). This evidences the 
much stronger phonon scattering strength induced by B$_{\text{C}}$ defects.
Furthermore, Fig.~\ref{fig:kappaconc} shows the dependence of $\kappa$ on impurity concentration, 
and confirms the anomalously large scattering from B$_\text{C}$ defects. Such a behavior of 
the B$_\text{C}$ defects seems the only possible explanation of the very low $\kappa$ (42~W/m/K and 19~W/m/K at 300~K) measured for 
the codoped samples. Otherwise, to achieve the same low 
thermal conductivities by N-doping alone would require the N$_{\text{C}}$ concentrations 
in excess of 20\%, which are obviously unrealistic.

\begin{figure}[t]
 \centering
  \includegraphics[width=9cm]{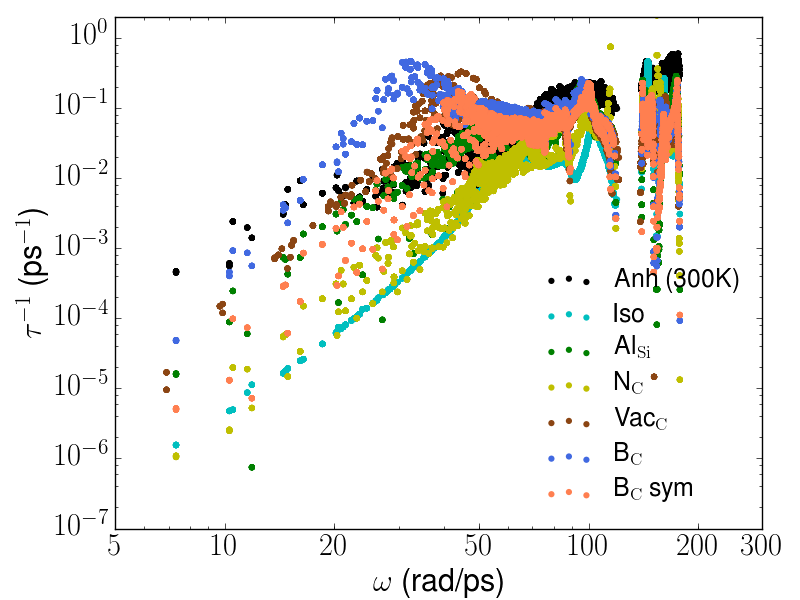} 
  \caption{Scattering rates of phonons from  different 
  defects (with concentration of 10$^{20}$ cm$^{-3}$), isotopes and 
  phonon-phonon interaction at 300~K. 
  }
  \label{fig:scat_rates} 
\end{figure}

A look at the phonon scattering rates in Fig.~\ref{fig:scat_rates} provides more clues 
for boron's atypical behavior. Scattering 
by substitutional impurities comes from two sources: the mass difference between the impurity atom and the one it replaces, and 
the local change in the interatomic force constants (IFCs) around the impurity in response to structural relaxation and modified chemical 
environment. In cases like Ge impurities in Si, the mass term dominates because Si and Ge have similar chemical bonding properties 
but very different masses \cite{Kundu_PRB2011, Garg_PRL2011}.
The other extreme is the case of vacancies, where the host atom is absent and the only contribution comes from IFC differences. The 
B-induced scattering rates in Fig.~\ref{fig:scat_rates}  are a striking two orders of magnitude larger than the N-induced ones, 
and they are comparable to those of the C vacancy. This is surprising because the absolute value of the mass difference between B and C is nearly the same as between N and C. A calculation including only mass difference, and neglecting IFC differences does indeed predict very similar scattering rates for B and N impurities (Fig. S1 of supplementary material.) Therefore the large scattering rates for B must come from the 
different changes in IFCs. However, scattering by Al substitutionals is not nearly as large as that of B, despite both species belonging to  
the same column in the periodic table, which might lead one to expect similar chemical bonding properties. 

\begin{figure}[t]
  \centering
  \includegraphics[width=9cm]{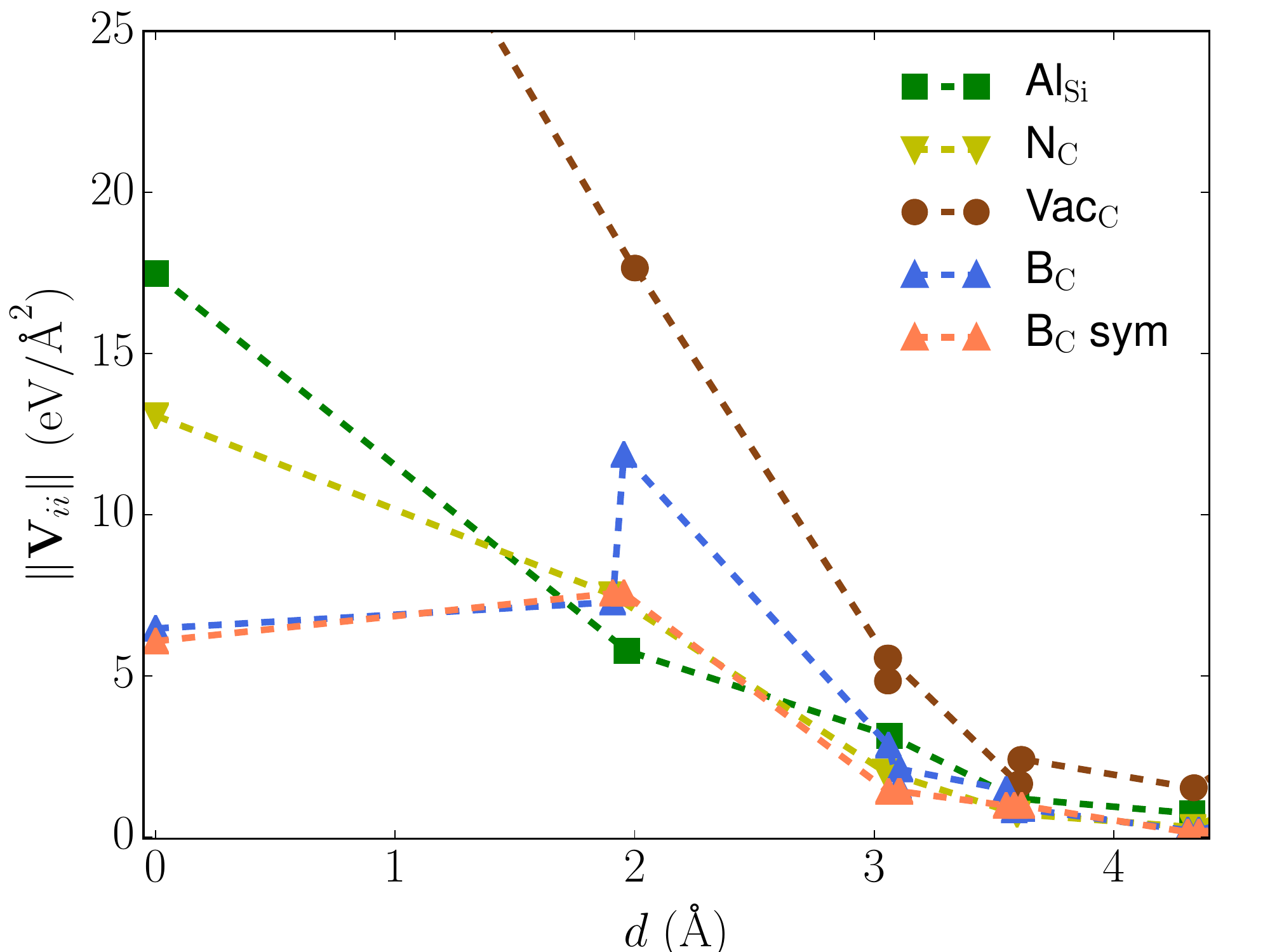}
  \caption{Norm of the changes in the on-site IFC submatrices of the atoms ($i$) around the defects with 
  respect to the perfect 3C-SiC structure as a function of the distance of the atoms from the defect centre ($d$).}
  \label{fig:fc_dif} 
\end{figure}

Fig. \ref{fig:fc_dif} shows the norm of the changes in the on-site IFC submatrices. Since translation 
symmetry imposes a set of linear constraints on each row of the IFC matrix, the change in the onsite terms 
probes the overall change in the IFCs related to each atom. The results shown in the figure confirm that the largest 
changes happen in the case of the vacancy. On the other hand, the B$_{\text{C}}$ defect shows a rather large asymmetry 
in the first-neighbor shell which may explain the increased scattering. To be able to analyze the contribution to 
scattering from this asymmetry, we define a symmetric perturbation matrix $\mathbf{V}_{sym}$ containing only the 
part of the total perturbation matrix $\mathbf{V}$ (see details in supplementary material) 
calculated by averaging $\mathbf{V}$ over the four space-group operations that just permute the four nearest-neighbors of the defect centre. 
Hence, $\mathbf{V}$ 
can be split as a sum of this symmetric term plus an asymmetric term containing the remainder of the total 
perturbation: $\mathbf{V}=\mathbf{V}_{sym}+\mathbf{V}_{asym}$. 
Figures \ref{fig:kappaconc} and \ref{fig:scat_rates} show the variation of $\kappa$ with defect concentration 
and the phonon-defect scattering rates induced by the artificial defect represented by $\mathbf{V}_{sym}$. 
Likewise, the norm of the changes in the on-site IFC submatrices 
due to $\mathbf{V}_{sym}$ is shown in Fig. \ref{fig:fc_dif}. It can be seen that the symmetrization process does 
indeed bring the values more in line with those corresponding to the symmetric perturbations  N$_{\text{C}}$ and 
Al$_{\text{Si}}$. Accordingly, the symmetrized B$_{\text{C}}$ perturbation yields thermal conductivities very 
similar to those from Al$_{\text{Si}}$ impurities, and much larger than those for the same concentration of 
true B$_{\text{C}}$ impurities, as shown in Fig. \ref{fig:kappaconc}.

The asymmetry in IFCs is caused by an asymmetry in the relaxed structure of the B$_{\text{C}}$ defect. While both N and 
Al lead to tetragonally symmetric relaxations around the impurity, the relaxation around B breaks the symmetry 
and the impurity gets closer to three of its four nearest-neighbors  (Fig. S2 in the supplementary material; see 
also Ref.~\onlinecite{Petrenko_prb2016}.) The asymmetrical relaxed structure of the B$_{\text{C}}$ defect is a 
manifestation of the complex chemistry of boron \cite{Oganov_jsm2009}. Some insight into the different chemical 
behaviors of N and B with respect to Si can be gained from a look at their binary compounds. While the most 
stable silicon nitride is Si$_3$N$_4$, with a relatively simple structure, silicon borides have stoichiometries 
from SiB$_3$ to SiB$_{40}$ and very complex unit cells with hundreds of atoms \cite{Vlasse_jssc1986, Aselage_jmr1998}. 
The stoichiometry and the structure of silicon borides suggest an affinity of B for low coordination with Si. 
Thus, relaxing of B defect atom away from one of its four neighboring atoms in 3C-SiC can be interpreted as its 
movement towards lower coordination.

Although the above reasoning shows that symmetry breaking affects the perturbation for 
B$_{\text{C}}$ impurities, this does not yet explain its exceptionally large 
effect on $\kappa$.
To understand it, let us take another look at the scattering rates in Fig.~\ref{fig:scat_rates}.
The scattering rates for B$_{\text{C}}$ display a prominent peak at about 33 rad/ps. A similarly large peak is observed for the vacancy at 
slightly higher frequency. Such peak is notably smaller or almost absent for the Al$_{\text{Si}}$ and N$_{\text{C}}$ defects. 
The rates for the symmetrized B$_{\text{C}}$ impurity are also shown (B$_{\text{C}}$ sym), displaying much smaller values and a much lower peak 
than the original B$_{\text{C}}$ perturbation.
This peak is the signature of resonant scattering. Resonances are quasi-bound states in the continuum of propagating 
states. They are infinitely extended, but have a large probability around the localized perturbation that causes them. They manifest 
themselves as peaks in the phonon scattering cross section ($\sigma$) of the scatterer near the frequency of the resonance. 
The Green's function approach that we use to compute defect scattering rates accounts for the full scattering to all orders, so 
it perfectly captures the effect of resonances, which would not appear in truncated perturbative approaches.

\begin{figure}[t]
 \centering
  \includegraphics[width=8.5cm]{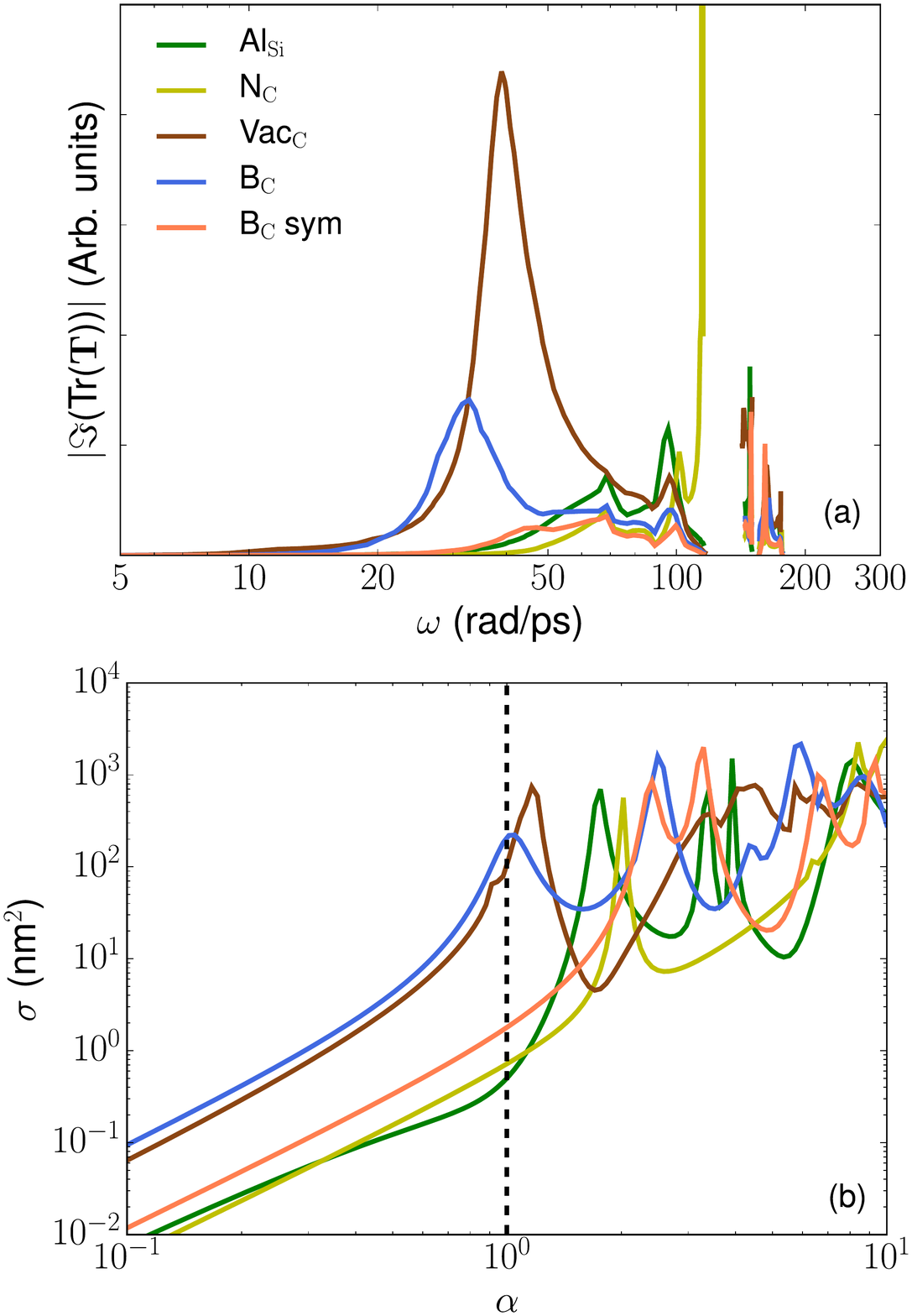} 
  \caption{(a) Trace of the imaginary part of the scattering T matrix for different impurities, showing a resonance for B and vacancy. (b) Scattering cross section ($\sigma$) for the LA mode at an angular frequency of 33 rad/ps, as a function of scattering strength for different impurities.}
  \label{fig:tmatrix} 
\end{figure}

A direct way to identify resonant scattering 
is via the scattering $\mathbf{T}$ matrix (see the expression in supplementary), 
which is fundamentally related to the cross section \cite{Mingo_PRB2010, Katcho_PRB2014}. 
Prominent peaks in the imaginary part of the trace of the scattering $\mathbf{T}$ matrix, 
shown in Fig.~\ref{fig:tmatrix}(a), correspond to resonances. In this way they are much easier to 
identify than from the cloud of points in the $\mathbf{q}$-point dependent scattering rate plots. The figure clearly displays 
large resonances for B$_{\text{C}}$ and C vacancies (Vac$_{\text{C}}$), 
but neither for the symmetrized B$_{\text{C}}$ nor for N$_{\text{C}}$ and Al$_{\text{Si}}$.

According to scattering theory, as the intensity of the localized perturbation grows stronger, increasingly sharper resonances start to develop at the lower frequency spectrum \cite{Economou_1983}.
In the present case, this would imply that resonances might show up for impurities other than B$_{\text{C}}$ or C vacancies, 
if only their perturbations were stronger. To verify this, we have artificially multiplied 
the perturbation by a constant factor $\alpha$, and re-calculated $\sigma$ for a longitudinal 
acoustic (LA) phonon of frequency 33 rad/ps (the resonant frequency for B$_{\text{C}}$ defect), as a function of $\alpha$. The value $\alpha=1$ 
corresponds to the true perturbation. Fig.~\ref{fig:tmatrix}(b) confirms that for sufficiently large $\alpha$, every 
impurity is able to develop resonant scattering. 
Fig.~\ref{fig:tmatrix}(b) suggests the qualitative concept of an effective native ``strength'' of the defect: the 
B$_{\text{C}}$ and C vacancy defects are associated to a larger strength, manifested as a larger cross section 
at small $\alpha$ and an earlier onset of the resonance with increase in $\alpha$. This “strength” is the 
result of a complex interplay involving the magnitude of the different changes in IFCs and their symmetry. 
In the B$_{\text{C}}$ case, the broken symmetry leads to a visibly enhanced strength with respect to the symmetrized 
case, seen in Fig.~\ref{fig:tmatrix}(b), 
despite the overall change in the  perturbation matrices in Fig 4, not being drastic.

In conclusion, B$_\text{C}$ substitutional impurities in cubic SiC scatter phonons much more strongly than N$_\text{C}$ or Al$_\text{Si}$ impurities, 
and comparably to C vacancies. For N impurities to reduce $\kappa$ as much as B impurities do, 
the concentration of the former has to be roughly 30 times larger than that of the latter. This could not have been guessed from 
mass difference considerations. It emerges from \textit{ab-initio} calculations, which are clearly supported by 
previous experimental measurements. This striking behavior is caused by the symmetry 
breaking around the B impurity upon relaxation, which leads to strong resonant scattering of acoustic phonons in the angular 
frequency range between 33 and 50 rad/ps. If only the symmetric part of the perturbation is kept, the resonance disappears 
and the calculated effect becomes similar to that of Al substitutionals. Furthermore, the calculations presented suggest that single-crystalline, 
defect-free cubic SiC can achieve a thermal conductivity about 1.7 times higher than the largest reported measured value on 
this polytype, and above than those of other polytypes. 

The results shown here point at a general phenomenon, whereby lattice distortions induced by 
symmetry breaking impurities may develop into a phonon resonance and lead to enhanced scattering. This should be relevant to many rapidly evolving technologies. 
In complex semiconductors there will usually be several possible doping scenarios for reaching a desired carrier 
concentration, and the sensitivity of the phonon scattering to the specific defect thus opens up an important new route in semiconductor design.

We acknowledge support from the Air Force Office of Scientific Research, USAF under award 
no. FA9550615-1-0187 DEF, and the European Union's Horizon 2020 Research and Innovation Programme, grant number 645776 (ALMA). We thank David Broido 
for valuable discussions. 


\begin{thebibliography}{34}
\expandafter\ifx\csname natexlab\endcsname\relax\def\natexlab#1{#1}\fi
\expandafter\ifx\csname bibnamefont\endcsname\relax
  \def\bibnamefont#1{#1}\fi
\expandafter\ifx\csname bibfnamefont\endcsname\relax
  \def\bibfnamefont#1{#1}\fi
\expandafter\ifx\csname citenamefont\endcsname\relax
  \def\citenamefont#1{#1}\fi
\expandafter\ifx\csname url\endcsname\relax
  \def\url#1{\texttt{#1}}\fi
\expandafter\ifx\csname urlprefix\endcsname\relax\def\urlprefix{URL }\fi
\providecommand{\bibinfo}[2]{#2}
\providecommand{\eprint}[2][]{\url{#2}}

\bibitem[{\citenamefont{Eddy and Gaskill}(2009)}]{Eddy_Sci2009}
\bibinfo{author}{\bibfnamefont{J.~C.~R.} \bibnamefont{Eddy}} \bibnamefont{and}
  \bibinfo{author}{\bibfnamefont{D.~K.} \bibnamefont{Gaskill}},
  \bibinfo{journal}{Science} \textbf{\bibinfo{volume}{324}},
  \bibinfo{pages}{1398} (\bibinfo{year}{2009}).

\bibitem[{\citenamefont{Bhatnagar and Baliga}(1993)}]{Bhatnagar_ieee1993}
\bibinfo{author}{\bibfnamefont{M.}~\bibnamefont{Bhatnagar}} \bibnamefont{and}
  \bibinfo{author}{\bibfnamefont{B.~J.} \bibnamefont{Baliga}},
  \bibinfo{journal}{IEEE Transac. Elec. Dev.} \textbf{\bibinfo{volume}{40}},
  \bibinfo{pages}{645} (\bibinfo{year}{1993}).

\bibitem[{\citenamefont{Fr\'echette and Carraro}(2006)}]{Frechette_jacs2006}
\bibinfo{author}{\bibfnamefont{J.}~\bibnamefont{Fr\'echette}} \bibnamefont{and}
  \bibinfo{author}{\bibfnamefont{C.}~\bibnamefont{Carraro}},
  \bibinfo{journal}{J. Am. Chem. Soc.} \textbf{\bibinfo{volume}{128}},
  \bibinfo{pages}{14774} (\bibinfo{year}{2006}).

\bibitem[{\citenamefont{Hao et~al.}(2016)\citenamefont{Hao, Guo, Pan, Chen,
  Jiao, Yang, and Guo}}]{Hao_jacs2016}
\bibinfo{author}{\bibfnamefont{C.}~\bibnamefont{Hao}},
  \bibinfo{author}{\bibfnamefont{X.}~\bibnamefont{Guo}},
  \bibinfo{author}{\bibfnamefont{Y.}~\bibnamefont{Pan}},
  \bibinfo{author}{\bibfnamefont{S.}~\bibnamefont{Chen}},
  \bibinfo{author}{\bibfnamefont{Z.}~\bibnamefont{Jiao}},
  \bibinfo{author}{\bibfnamefont{H.}~\bibnamefont{Yang}}, \bibnamefont{and}
  \bibinfo{author}{\bibfnamefont{X.}~\bibnamefont{Guo}}, \bibinfo{journal}{J.
  Am. Chem. Soc.} \textbf{\bibinfo{volume}{138}}, \bibinfo{pages}{9361}
  (\bibinfo{year}{2016}).

\bibitem[{\citenamefont{Saddow}(2012)}]{Saddow_2012}
\bibinfo{author}{\bibfnamefont{S.~E.} \bibnamefont{Saddow}},
  \emph{\bibinfo{title}{Silicon Carbide Biotechnology}}
  (\bibinfo{publisher}{Elsiver: Amsterdam}, \bibinfo{year}{2012}).

\bibitem[{\citenamefont{Hillenbrand et~al.}(2002)\citenamefont{Hillenbrand,
  Taubner, and Keilmann}}]{Hillenbrand_Nature2002}
\bibinfo{author}{\bibfnamefont{R.}~\bibnamefont{Hillenbrand}},
  \bibinfo{author}{\bibfnamefont{T.}~\bibnamefont{Taubner}}, \bibnamefont{and}
  \bibinfo{author}{\bibfnamefont{F.}~\bibnamefont{Keilmann}},
  \bibinfo{journal}{Nature} \textbf{\bibinfo{volume}{418}},
  \bibinfo{pages}{159} (\bibinfo{year}{2002}).

\bibitem[{\citenamefont{Syvajarvi et~al.}(2016)\citenamefont{Syvajarvi, Ma,
  Jokubavicius, Galeckas, Sun, Liu, Jansson, Wellmann, Linnarsson, Runde
  et~al.}}]{Mikael_SEMSC2016}
\bibinfo{author}{\bibfnamefont{M.}~\bibnamefont{Syvajarvi}},
  \bibinfo{author}{\bibfnamefont{Q.}~\bibnamefont{Ma}},
  \bibinfo{author}{\bibfnamefont{V.}~\bibnamefont{Jokubavicius}},
  \bibinfo{author}{\bibfnamefont{A.}~\bibnamefont{Galeckas}},
  \bibinfo{author}{\bibfnamefont{J.}~\bibnamefont{Sun}},
  \bibinfo{author}{\bibfnamefont{X.}~\bibnamefont{Liu}},
  \bibinfo{author}{\bibfnamefont{M.}~\bibnamefont{Jansson}},
  \bibinfo{author}{\bibfnamefont{P.}~\bibnamefont{Wellmann}},
  \bibinfo{author}{\bibfnamefont{M.}~\bibnamefont{Linnarsson}},
  \bibinfo{author}{\bibfnamefont{P.}~\bibnamefont{Runde}},
  \bibnamefont{et~al.}, \bibinfo{journal}{Solar Ene. Mater. and Solar Cel.}
  \textbf{\bibinfo{volume}{145, Part 2}}, \bibinfo{pages}{104}
  (\bibinfo{year}{2016}).

\bibitem[{\citenamefont{Beaucarne et~al.}(2002)\citenamefont{Beaucarne, Brown,
  Keevers, Corkish, and Green}}]{Beaucarne_PPRA2002}
\bibinfo{author}{\bibfnamefont{G.}~\bibnamefont{Beaucarne}},
  \bibinfo{author}{\bibfnamefont{A.~S.} \bibnamefont{Brown}},
  \bibinfo{author}{\bibfnamefont{M.~J.} \bibnamefont{Keevers}},
  \bibinfo{author}{\bibfnamefont{R.}~\bibnamefont{Corkish}}, \bibnamefont{and}
  \bibinfo{author}{\bibfnamefont{M.~A.} \bibnamefont{Green}},
  \bibinfo{journal}{Prog. Photovolt: Res. Appl.} \textbf{\bibinfo{volume}{10}},
  \bibinfo{pages}{345} (\bibinfo{year}{2002}).

\bibitem[{\citenamefont{Cicero et~al.}(2004)\citenamefont{Cicero, Catellani,
  and Galli}}]{Cicero_PRL2004}
\bibinfo{author}{\bibfnamefont{G.}~\bibnamefont{Cicero}},
  \bibinfo{author}{\bibfnamefont{A.}~\bibnamefont{Catellani}},
  \bibnamefont{and} \bibinfo{author}{\bibfnamefont{G.}~\bibnamefont{Galli}},
  \bibinfo{journal}{Phys. Rev. Lett.} \textbf{\bibinfo{volume}{93}},
  \bibinfo{pages}{016102} (\bibinfo{year}{2004}).

\bibitem[{\citenamefont{Harris~(ed.)}(1995)}]{Harris_1995}
\bibinfo{author}{\bibfnamefont{G.~L.} \bibnamefont{Harris~(ed.)}}, in
  \emph{\bibinfo{booktitle}{Properties of Silicon Carbide}}
  (\bibinfo{publisher}{ISPEC, London}, \bibinfo{year}{1995}).

\bibitem[{\citenamefont{Zheludev}(2007)}]{Zheludev_2007}
\bibinfo{author}{\bibfnamefont{N.}~\bibnamefont{Zheludev}},
  \bibinfo{journal}{Nat. Phot.} \textbf{\bibinfo{volume}{1}},
  \bibinfo{pages}{189} (\bibinfo{year}{2007}).

\bibitem[{\citenamefont{W.~C.~Lien et~al.}(2014)\citenamefont{W.~C.~Lien,
  Damrongplasit, Paredes, G., Liu, and Pisano}}]{Lien_ieee2014}
\bibinfo{author}{\bibfnamefont{W.~C.} \bibnamefont{W.~C.~Lien}},
  \bibinfo{author}{\bibfnamefont{N.}~\bibnamefont{Damrongplasit}},
  \bibinfo{author}{\bibfnamefont{J.~H.} \bibnamefont{Paredes}},
  \bibinfo{author}{\bibfnamefont{S.~D.} \bibnamefont{G.}},
  \bibinfo{author}{\bibfnamefont{T.~J.~K.} \bibnamefont{Liu}},
  \bibnamefont{and} \bibinfo{author}{\bibfnamefont{A.~P.}
  \bibnamefont{Pisano}}, \bibinfo{journal}{IEEE J. Elec. Dev. Soc.}
  \textbf{\bibinfo{volume}{2}}, \bibinfo{pages}{164} (\bibinfo{year}{2014}).

\bibitem[{\citenamefont{Verucchi et~al.}(2012)\citenamefont{Verucchi, Aversa,
  Nardi, Taioli, a~Beccara, Alf\`e, Nasi, Rossi, Salviati, and
  Iannotta}}]{Veruchhi_jacs2012}
\bibinfo{author}{\bibfnamefont{R.}~\bibnamefont{Verucchi}},
  \bibinfo{author}{\bibfnamefont{L.}~\bibnamefont{Aversa}},
  \bibinfo{author}{\bibfnamefont{M.~V.} \bibnamefont{Nardi}},
  \bibinfo{author}{\bibfnamefont{S.}~\bibnamefont{Taioli}},
  \bibinfo{author}{\bibfnamefont{S.}~\bibnamefont{a~Beccara}},
  \bibinfo{author}{\bibfnamefont{D.}~\bibnamefont{Alf\`e}},
  \bibinfo{author}{\bibfnamefont{L.}~\bibnamefont{Nasi}},
  \bibinfo{author}{\bibfnamefont{F.}~\bibnamefont{Rossi}},
  \bibinfo{author}{\bibfnamefont{G.}~\bibnamefont{Salviati}}, \bibnamefont{and}
  \bibinfo{author}{\bibfnamefont{S.}~\bibnamefont{Iannotta}},
  \bibinfo{journal}{J. Am. Chem. Soc.} \textbf{\bibinfo{volume}{134}},
  \bibinfo{pages}{17400} (\bibinfo{year}{2012}).

\bibitem[{\citenamefont{Morelli et~al.}(1993)\citenamefont{Morelli, Heremans,
  Beetz, Woo, Harris, and Taylor}}]{Morelli_1993}
\bibinfo{author}{\bibfnamefont{D.}~\bibnamefont{Morelli}},
  \bibinfo{author}{\bibfnamefont{J.}~\bibnamefont{Heremans}},
  \bibinfo{author}{\bibfnamefont{C.}~\bibnamefont{Beetz}},
  \bibinfo{author}{\bibfnamefont{W.~S.} \bibnamefont{Woo}},
  \bibinfo{author}{\bibfnamefont{G.~L.} \bibnamefont{Harris}},
  \bibnamefont{and} \bibinfo{author}{\bibfnamefont{C.}~\bibnamefont{Taylor}},
  \bibinfo{journal}{Instit. Phys. Conf. Ser.} \textbf{\bibinfo{volume}{N137}},
  \bibinfo{pages}{313} (\bibinfo{year}{1993}).

\bibitem[{\citenamefont{Ivanova et~al.}(2006)\citenamefont{Ivanova,
  Aleksandrov, and Demakov}}]{Ivanova_im2006}
\bibinfo{author}{\bibfnamefont{L.~M.} \bibnamefont{Ivanova}},
  \bibinfo{author}{\bibfnamefont{P.~A.} \bibnamefont{Aleksandrov}},
  \bibnamefont{and} \bibinfo{author}{\bibfnamefont{K.~D.}
  \bibnamefont{Demakov}}, \bibinfo{journal}{Inorg. Mater.}
  \textbf{\bibinfo{volume}{42}}, \bibinfo{pages}{1205} (\bibinfo{year}{2006}).

\bibitem[{\citenamefont{Li et~al.}(2014)\citenamefont{Li, Carrete, Katcho, and
  Mingo}}]{Li_CPC2014}
\bibinfo{author}{\bibfnamefont{W.}~\bibnamefont{Li}},
  \bibinfo{author}{\bibfnamefont{J.}~\bibnamefont{Carrete}},
  \bibinfo{author}{\bibfnamefont{N.~A.} \bibnamefont{Katcho}},
  \bibnamefont{and} \bibinfo{author}{\bibfnamefont{N.}~\bibnamefont{Mingo}},
  \bibinfo{journal}{Comp. Phys. Comm.} \textbf{\bibinfo{volume}{185}},
  \bibinfo{pages}{1747 } (\bibinfo{year}{2014}).

\bibitem[{\citenamefont{Katre and Madsen}(2016)}]{Katre_PRB2016}
\bibinfo{author}{\bibfnamefont{A.}~\bibnamefont{Katre}} \bibnamefont{and}
  \bibinfo{author}{\bibfnamefont{G.~K.~H.} \bibnamefont{Madsen}},
  \bibinfo{journal}{Phys. Rev. B} \textbf{\bibinfo{volume}{93}},
  \bibinfo{pages}{155203} (\bibinfo{year}{2016}).

\bibitem[{\citenamefont{{The ALMA project developers}}()}]{ALMA_BTE}
\bibinfo{author}{\bibnamefont{{The ALMA project developers}}},
  \emph{\bibinfo{title}{All-scale predictive design of heat management material
  structures}}, \bibinfo{howpublished}{\url{http://www.almabte.eu/}}.

\bibitem[{\citenamefont{{Ioffe Institute}}()}]{ioffe_sic}
\bibinfo{author}{\bibnamefont{{Ioffe Institute}}},
  \emph{\bibinfo{title}{Thermal properties of silicon carbide}},
  \bibinfo{howpublished}{\url{http://www.ioffe.ru/SVA/NSM/Semicond/SiC/thermal.html}}.

\bibitem[{\citenamefont{Gupta and Ballato}(2006)}]{Gupta_2006}
\bibinfo{author}{\bibfnamefont{M.~C.} \bibnamefont{Gupta}} \bibnamefont{and}
  \bibinfo{author}{\bibfnamefont{J.}~\bibnamefont{Ballato}},
  \emph{\bibinfo{title}{The Handbook of Photonics, Second Edition}}
  (\bibinfo{publisher}{CRC Press}, \bibinfo{year}{2006}).

\bibitem[{\citenamefont{Smith et~al.}(1999)\citenamefont{Smith, Evwaraye,
  Mitchel, and Capano}}]{Smith_jem1999}
\bibinfo{author}{\bibfnamefont{S.~R.} \bibnamefont{Smith}},
  \bibinfo{author}{\bibfnamefont{A.~O.} \bibnamefont{Evwaraye}},
  \bibinfo{author}{\bibfnamefont{W.~C.} \bibnamefont{Mitchel}},
  \bibnamefont{and} \bibinfo{author}{\bibfnamefont{M.~A.}
  \bibnamefont{Capano}}, \bibinfo{journal}{J. Elec. Mater.}
  \textbf{\bibinfo{volume}{28}}, \bibinfo{pages}{190} (\bibinfo{year}{1999}).

\bibitem[{\citenamefont{Isoya et~al.}(2003)\citenamefont{Isoya, Ohshima,
  Morishita, Kamiya, Itoh, and Yamasaki}}]{Isoya_pbcm2003}
\bibinfo{author}{\bibfnamefont{J.}~\bibnamefont{Isoya}},
  \bibinfo{author}{\bibfnamefont{T.}~\bibnamefont{Ohshima}},
  \bibinfo{author}{\bibfnamefont{N.}~\bibnamefont{Morishita}},
  \bibinfo{author}{\bibfnamefont{T.}~\bibnamefont{Kamiya}},
  \bibinfo{author}{\bibfnamefont{H.}~\bibnamefont{Itoh}}, \bibnamefont{and}
  \bibinfo{author}{\bibfnamefont{S.}~\bibnamefont{Yamasaki}},
  \bibinfo{journal}{Phys. B: Cond. Mat.} \textbf{\bibinfo{volume}{340–342}},
  \bibinfo{pages}{903 } (\bibinfo{year}{2003}).

\bibitem[{\citenamefont{Fukumoto}(1996)}]{Fukumoto_prb1996}
\bibinfo{author}{\bibfnamefont{A.}~\bibnamefont{Fukumoto}},
  \bibinfo{journal}{Phys. Rev. B} \textbf{\bibinfo{volume}{53}},
  \bibinfo{pages}{4458} (\bibinfo{year}{1996}).

\bibitem[{\citenamefont{Bechstedt et~al.}(2001)\citenamefont{Bechstedt, Fissel,
  Furthm\"uller, Grossner, and Zywietz}}]{Bechstedt_jpcm2001}
\bibinfo{author}{\bibfnamefont{F.}~\bibnamefont{Bechstedt}},
  \bibinfo{author}{\bibfnamefont{A.}~\bibnamefont{Fissel}},
  \bibinfo{author}{\bibfnamefont{J.}~\bibnamefont{Furthm\"uller}},
  \bibinfo{author}{\bibfnamefont{U.}~\bibnamefont{Grossner}}, \bibnamefont{and}
  \bibinfo{author}{\bibfnamefont{A.}~\bibnamefont{Zywietz}},
  \bibinfo{journal}{J. Phys.: Cond. Mat.} \textbf{\bibinfo{volume}{13}},
  \bibinfo{pages}{9027} (\bibinfo{year}{2001}).

\bibitem[{\citenamefont{Petrenko and Petrenko}(2016)}]{Petrenko_prb2016}
\bibinfo{author}{\bibfnamefont{T.~T.} \bibnamefont{Petrenko}} \bibnamefont{and}
  \bibinfo{author}{\bibfnamefont{T.~L.} \bibnamefont{Petrenko}},
  \bibinfo{journal}{Phys. Rev. B} \textbf{\bibinfo{volume}{93}},
  \bibinfo{pages}{165203} (\bibinfo{year}{2016}).

\bibitem[{\citenamefont{Oda et~al.}({2013})\citenamefont{Oda, Zhang, and
  Weber}}]{Oda_JCP13}
\bibinfo{author}{\bibfnamefont{T.}~\bibnamefont{Oda}},
  \bibinfo{author}{\bibfnamefont{Y.}~\bibnamefont{Zhang}}, \bibnamefont{and}
  \bibinfo{author}{\bibfnamefont{W.~J.} \bibnamefont{Weber}},
  \bibinfo{journal}{J. Chem. Phys.} \textbf{\bibinfo{volume}{{139}}}
  (\bibinfo{year}{{2013}}).

\bibitem[{\citenamefont{Kundu et~al.}(2011)\citenamefont{Kundu, Mingo, Broido,
  and Stewart}}]{Kundu_PRB2011}
\bibinfo{author}{\bibfnamefont{A.}~\bibnamefont{Kundu}},
  \bibinfo{author}{\bibfnamefont{N.}~\bibnamefont{Mingo}},
  \bibinfo{author}{\bibfnamefont{D.~A.} \bibnamefont{Broido}},
  \bibnamefont{and} \bibinfo{author}{\bibfnamefont{D.~A.}
  \bibnamefont{Stewart}}, \bibinfo{journal}{Phys. Rev. B}
  \textbf{\bibinfo{volume}{84}}, \bibinfo{pages}{125426}
  (\bibinfo{year}{2011}).

\bibitem[{\citenamefont{Garg et~al.}(2011)\citenamefont{Garg, Bonini, Kozinsky,
  and Marzari}}]{Garg_PRL2011}
\bibinfo{author}{\bibfnamefont{J.}~\bibnamefont{Garg}},
  \bibinfo{author}{\bibfnamefont{N.}~\bibnamefont{Bonini}},
  \bibinfo{author}{\bibfnamefont{B.}~\bibnamefont{Kozinsky}}, \bibnamefont{and}
  \bibinfo{author}{\bibfnamefont{N.}~\bibnamefont{Marzari}},
  \bibinfo{journal}{Phys. Rev. Lett.} \textbf{\bibinfo{volume}{106}},
  \bibinfo{pages}{045901} (\bibinfo{year}{2011}).

\bibitem[{\citenamefont{Oganov and Solozhenko}(2009)}]{Oganov_jsm2009}
\bibinfo{author}{\bibfnamefont{A.~R.} \bibnamefont{Oganov}} \bibnamefont{and}
  \bibinfo{author}{\bibfnamefont{V.~L.} \bibnamefont{Solozhenko}},
  \bibinfo{journal}{J. Superhard Mater.} \textbf{\bibinfo{volume}{31}},
  \bibinfo{pages}{285} (\bibinfo{year}{2009}).

\bibitem[{\citenamefont{Vlasse et~al.}(1986)\citenamefont{Vlasse, Slack,
  Garbauskas, Kasper, and Viala}}]{Vlasse_jssc1986}
\bibinfo{author}{\bibfnamefont{M.}~\bibnamefont{Vlasse}},
  \bibinfo{author}{\bibfnamefont{G.~A.} \bibnamefont{Slack}},
  \bibinfo{author}{\bibfnamefont{M.}~\bibnamefont{Garbauskas}},
  \bibinfo{author}{\bibfnamefont{J.~S.} \bibnamefont{Kasper}},
  \bibnamefont{and} \bibinfo{author}{\bibfnamefont{J.~C.} \bibnamefont{Viala}},
  \bibinfo{journal}{J. Solid Stat. Chem.} \textbf{\bibinfo{volume}{63}},
  \bibinfo{pages}{31 } (\bibinfo{year}{1986}).

\bibitem[{\citenamefont{Aselage}(1998)}]{Aselage_jmr1998}
\bibinfo{author}{\bibfnamefont{T.~L.} \bibnamefont{Aselage}},
  \bibinfo{journal}{J. Mater. Research} \textbf{\bibinfo{volume}{13}},
  \bibinfo{pages}{1786–1794} (\bibinfo{year}{1998}).

\bibitem[{\citenamefont{Mingo et~al.}(2010)\citenamefont{Mingo, Esfarjani,
  Broido, and Stewart}}]{Mingo_PRB2010}
\bibinfo{author}{\bibfnamefont{N.}~\bibnamefont{Mingo}},
  \bibinfo{author}{\bibfnamefont{K.}~\bibnamefont{Esfarjani}},
  \bibinfo{author}{\bibfnamefont{D.~A.} \bibnamefont{Broido}},
  \bibnamefont{and} \bibinfo{author}{\bibfnamefont{D.~A.}
  \bibnamefont{Stewart}}, \bibinfo{journal}{Phys. Rev. B}
  \textbf{\bibinfo{volume}{81}}, \bibinfo{pages}{045408}
  (\bibinfo{year}{2010}).

\bibitem[{\citenamefont{Katcho et~al.}(2014)\citenamefont{Katcho, Carrete, Li,
  and Mingo}}]{Katcho_PRB2014}
\bibinfo{author}{\bibfnamefont{N.~A.} \bibnamefont{Katcho}},
  \bibinfo{author}{\bibfnamefont{J.}~\bibnamefont{Carrete}},
  \bibinfo{author}{\bibfnamefont{W.}~\bibnamefont{Li}}, \bibnamefont{and}
  \bibinfo{author}{\bibfnamefont{N.}~\bibnamefont{Mingo}},
  \bibinfo{journal}{Phys. Rev. B} \textbf{\bibinfo{volume}{90}},
  \bibinfo{pages}{094117} (\bibinfo{year}{2014}).

\bibitem[{\citenamefont{Economou}(1983)}]{Economou_1983}
\bibinfo{author}{\bibfnamefont{E.}~\bibnamefont{Economou}}, in
  \emph{\bibinfo{booktitle}{Green's function in quantum Physics}}, edited by
  \bibinfo{editor}{\bibfnamefont{M.}~\bibnamefont{Cardona}},
  \bibinfo{editor}{\bibfnamefont{P.}~\bibnamefont{Fulde}}, \bibnamefont{and}
  \bibinfo{editor}{\bibfnamefont{H.~J.} \bibnamefont{Queisser}}
  (\bibinfo{publisher}{Springer New York}, \bibinfo{year}{1983}), pp.
  \bibinfo{pages}{108--109}, \bibinfo{edition}{{II}} ed.

\end{thebibliography}


\end{document}